\documentclass[apsrev4,pra,superscriptaddress,notitlepage,nofootinbib]{revtex4-1} %% REVTeX 4.0
\usepackage[colorlinks=true,allcolors=blue]{hyperref}
\usepackage{amsmath,amsfonts,amssymb}
\usepackage{graphicx,xcolor}
\usepackage[version=4]{mhchem}
\begin{document} 
\title{Commissioning measurements for a very cold neutron interferometer based on nanodiamond-polymer composite gratings}
%%%%%%%%%%%%%%%%%%%%
\author{Roxana H. Ackermann}
\author{Sonja Falmbigl}
\affiliation{Faculty of Physics, University of Vienna, Wien, Austria}
\author{Elhoucine Hadden}\thanks{now at: Department of Physics and Astronomy, Uppsala University, Sweden}
\affiliation{Institut Laue-Langevin, Grenoble, France}
\author{Alexia Dubois Leprou}
\affiliation{Institut Laue-Langevin, Grenoble, France}
\affiliation{Grenoble INP-Phelma, Université Grenoble Alpes, France}
\author{Hanno Filter-Pieler}
\affiliation{Institut Laue-Langevin, Grenoble, France}
\author{Tobias Jenke}
\affiliation{Institut Laue-Langevin, Grenoble, France}
\author{J\"urgen Klepp}
\affiliation{Faculty of Physics, University of Vienna, Wien, Austria}
\author{Christian Pruner}
\affiliation{Faculty of Natural and Life Sciences, Department of Chemistry and Physics of Materials, University of Salzburg, Austria}
\author{Yasuo Tomita}
\affiliation{Department of Engineering Science, University of Electro-Communications, Tokyo, Japan}
\author{Martin Fally}
\thanks{send correspondence to M.F.: martin.fally@univie.ac.at}
\affiliation{Faculty of Physics, University of Vienna, Wien, Austria}

%%%%%%%%%%%%%%%%%%%
\hypersetup{pdfauthor={Ackermann, Falmbigl, Hadden, Dubois Leprou, Filter-Pieler, Jenke, Klepp, Pruner, Tomita, Fally},pdftitle={Commissioning measurements for a very cold neutron interferometer based on nanodiamond-polymer composite gratings},pdfborder={0 0 0},urlcolor=blue}
%%%%%%%%%%%%%%%%%%%%%%%%%%%%%%%%%%%%%%%%%%%%%%%%%%%%%%%%%%%%%%%%%%%%%%%%%%%5
% Option to view page numbers
\pagestyle{empty} % change to \pagestyle{plain} for page numbers   
\setcounter{page}{1} % Set start page numbering at e.g. 301
% % FALLY ABBREV
\newcommand{\bcr}{[b_c\Delta\rho]_1}
%UNIVIE colors and abbreviations
\definecolor{ured}{RGB}{167,28,73}%A71C49
\definecolor{ublue}{RGB}{0,99,166}%0063A6
\definecolor{ugreen}{RGB}{148,193,84}%94C154
\definecolor{ugray}{RGB}{102,102,102}%666666
\definecolor{uorange}{RGB}{221,72,20}%DD4814
\definecolor{uyellow}{RGB}{246,168,0}%F6A800
\definecolor{umint}{RGB}{17,137,122}%11897A
% % % % % % % % % % % % % % % % % % % % % % 
\newcommand{\tcr}[1]{\textcolor{ured}{#1}}
\newcommand{\tcb}[1]{\textcolor{ublue}{#1}}
\newcommand{\yco}[1]{\textcolor{red}{#1}}
\newcommand{\eco}[1]{\textcolor{black}{#1}}
\newcommand{\hco}[1]{\textcolor{blue}{#1}}
\newcommand{\cco}[1]{\textcolor{yellow}{#1}}

\graphicspath{{/home/fallym4/MyConferences/26/Photonics_Europe_Strasbourg/fig/}}

\begin{abstract}
Over the past decade, holographic nanodiamond–polymer composite gratings have been developed and optimized as high‑efficiency diffractive elements for very cold neutrons (VCN), for use as mirrors and beam splitters in a triple‑Laue (LLL) interferometer. We report their optical characterization and, crucially, their neutron-optical performance, including diffraction efficiency and angular selectivity under VCN conditions. We further demonstrate their integration into a VCN interferometer. The layout of the interferometer and its first implementation at the beamline are described, highlighting practical considerations for long‑term operation. We discuss avenues for performance improvement, in particular grating fabrication refinements. These results establish nanodiamond–polymer composite gratings as viable components for VCN interferometry and pave a way toward precision neutron phase measurements in the very cold regime.
\end{abstract}

\maketitle
\section{INTRODUCTION}
\label{sec:intro}  % \label{} allows reference to this section
Interferometers for thermal neutrons based on a monolithic silicon single crystals in LLL-geometry have been established for more than half a century  \cite{Rauch-pla74}. Using such devices, 
%a vast number of experiments  dealing with aspects of both, the foundations of quantum mechanics as well as applied material sciences, were performed. 
a vast number of experiments addressing both fundamental aspects of quantum mechanics and applied materials science have been performed.
For an exhaustive overview see Refs.\, \cite{Rauch-15,Klepp-ptep14}. Recent results include measurements using weak measurements 
 \cite{Denkmayr-nc14,Denkmayr-pbcm18,Danner-sr24}, exploring the neutron angular orbital momentum  \cite{Sarenac-pnas19}, phase vortex lattices  \cite{Geerits-cp23} and cone beam interferometry  \cite{Sarenac-prr24} to name a few.

To extend the applicability of neutron interferometers to lower energy ranges, i.e., cold and very cold neutrons (VCN), other strategies using either surface-relief gratings  \cite{Ioffe-pla85}, multilayer mirrors  \cite{Funahashi-pra96} or artificial phase gratings were employed  \cite{Gruber-pla89,Eder-nima89,Zouw-nima00}.  
% Our approach, which is described here, uses holographically prepared volume phase gratings in nanodiamond-photopolymer composites (nDPC)
In the approach described here, holographically prepared volume phase gratings in nanodiamond-photopolymer composites (nDPC) are used, 
following procedures developed for all‑polymer grating interferometers as designed for cold neutrons  \cite{Schellhorn-pb97,Pruner-nima06}. Detailed information on the choice, the preparation and the general performance of the nDPC-gratings can be found in  Refs.\, \cite{Tomita-pra20,Hadden-apl24}. The basic setup method of the VCN interferometer including the crucial alignment strategy is described in Ref.\, \cite{Falmbigl-spie25}. 
% Here, we discuss the light and neutron optical performance of the actual interferometer gratings for VCN and describe 
Here, the light and neutron-optical performance of the actual interferometer gratings for VCN is discussed, and
their integration into the interferometer at the PF2-VCN beamline of the Institut Laue-Langevin, France, is described. 
Two setup and measurement campaigns  \cite{Klepp-InstitutLaue-LangevinILL25,Klepp-InstitutLaue-LangevinILL25a}
% 
% (\url{http://doi.org/10.5291/ILL-DATA.3-14-455}, \url{http://doi.org/10.5291/ILL-DATA.DIR-407}) 
% 
were conducted in May/June and September/October 2025 to commission the interferometer with prospective improvements to the environment, known to be of central importance  \cite{Huber-ewc19}, as well as to the gratings and their mutual alignment. The general setup design of the VCN interferometer was recently published  \cite{Falmbigl-spie25}, a detailed discussion can be found in Ref.  \cite{Ackermann-M26}. 
In what follows, the first attempt is discussed together with the reasoning behind the improvements to the VCN interferometer introduced in the second campaign, and possible future amendments are considered.

\section{Choice of the grating parameters: \\ grating period and thickness(es)}\label{sec:2}
Three nDPC-gratings $\text{G}_1,\text{G}_2,\text{G}_3,$ exhibiting the alignment process during holographic recording  \cite{Falmbigl-spie25} were prepared and investigated.
% \yco{mF2yT: the setup is shown in the reference cited here (our proc.-paper of last year)} 
The requirements for the ideal case in the Mach-Zehnder type setup are: $\text{G}_2$ should serve as a mirror (for both beam paths), the other two gratings as 50:50 beamsplitters. Any deviation from this ideal case will reduce the visibility of the interference signal and/or the total count rate at the detector. For holographically recorded gratings we have  \cite{Rauch-15} as usual
\newcommand{\ej}[2]{\eta^{(#1)}_{#2}} 
\begin{subequations}
\label{eq:1}
\begin{align}
I_0(\Delta\chi)&=a+c\cos(\Delta\chi)\label{eq:1a}\\
I_H(\Delta\chi)&=b-c\cos(\Delta\chi)\label{eq:1b}
\end{align}
\end{subequations}
for the intensity in the $0-$ and $H-$ beams with $\Delta\chi$, the phase difference between the interfering beams (see Fig.\,\ref{fig:VCN} bottom panel).
These constants ($a,b,c$) in contrast to other fully symmetric configurations  \cite{Rauch-15} with all identical gratings are given by:
\begin{subequations}\label{eq:I0H}
\begin{align}	
a&={\cal A}\ej{2}{1}\left[\ej{1}{0}\ej{3}{1}+\ej{1}{1}\ej{3}{0}\right]\\
b&={\cal A}\ej{2}{1}\left[\ej{1}{0}\ej{3}{0}+\ej{1}{1}\ej{3}{1}\right]\\
c&=2{\cal A}\ej{2}{1}\sqrt{\ej{1}{1}\ej{1}{0}\ej{3}{1}\ej{3}{0}}
\end{align}
\end{subequations}

with $\eta^{(j)}_m$ the $m$-th ($m=0,1$) order diffraction efficiency ("reflectivity") of grating $\text{G}_j$ and ${\cal A}$ the overall attenuation factor due to (in)coherent scattering, absorption and (negligible) optical reflection from surfaces  \cite{Boothroyd-20}. Moreover, in Eqs. (\ref{eq:I0H}) it is assumed that the beams transmitted through $\text{G}_2$ are blocked by proper slits whenever $\eta^{(2)}_1<1$, thus leading to a loss of neutrons without any influence on the interfering beams. Note that for standard triple Laue interferometers, using thermal neutrons, $\ej{j}{m} \leq 0.5$ for all diffraction elements because of averaging over rapidly oscillating Pendell\"osung due to the thickness of a few millimeters. This is \textbf{not} the case for our gratings with a few ten micrometers thickness.

For many experiments, the phase sensitivity of the interferometer basically depends on the area enclosed by the two spatially separated beams  \cite{Rauch-15}. Diffraction here is elastic and its efficiency is maximum when the following Bragg (or equivalently the Laue) condition is fulfilled:
\begin{equation}
2\Lambda\sin(\Theta)=m\lambda \label{eq:Bragg}
\end{equation}
with $\Lambda$ the grating period, $\Theta$ the angle of incidence, $\lambda$ the wavelength and $m\in\mathbb{Z}\backslash\{0\}$.
% , relating the wavelength, the grating period and the angle of incidence to account for momentum conservation, . 
Gratings prepared by visible light optical holography ($\lambda=\lambda_\text{L}\approx 500\,\text{nm}$), therefore, will result in grating periods of a few hundred nanometers, with shorter periods giving larger diffraction angles and finally a larger enclosed area between the beam paths. 
A target period of $\Lambda\approx 500\,\text{nm}$ was chosen, for which nDPC materials are known to work well \cite{Hadden-apl24}. 
VCN on the other hand have wavelengths of only a few nanometers ($\lambda_\text{N}=1 \ldots 10\,\text{nm}$), 2 to 3 orders of magnitude shorter than the spacing. 
Their actual wavelength distribution $S(\lambda_\text{N})$ at the PF2-VCN beamline depends on the exact position and the size of the entrance aperture as well as on the type of subsequent mirror that redirects the beam  \cite{Oda-nima17},
% \textcolor{uorange}{which paper to cite here?or Valentin's or Elhoucine's PhD?}, 
and therefore the properties of the gratings must be designed according to the expected profile of the wavelength distribution. One example is a rather broad wavelength distribution with a mean wavelength of about $\overline{\lambda}_\text{N}\approx 5.5\,\text{nm}$ and a width of $\Delta\lambda_\text{N}/\overline{\lambda}_\text{N} \approx 0.3$. Its shape can be approximated to an exponentially modified Gaussian (EMG) distribution  \cite{Blaickner-nima19,Hadden-apl24} (with three free parameters) or another three-parametric function  \cite{Abele-nima06}, without assumption of specific physical models. For such an assumed wavelength distribution and an overall interferometer length of $1500\,\text{mm}$ a beam separation of about $8\,\text{mm}$ at the mirror grating ($\text{G}_2$) is expected with an enclosed area between the beampaths of $60\,\text{cm}^2$.

Whenever the Bragg condition is violated, dephasing occurs and the diffraction efficiency decreases. The angular selectivity of the diffraction efficiency increases roughly proportional to the thickness $d$ of the grating. The diffraction efficiency itself is proportional to the square of the thickness \cite{Kogelnik-atj69}. A broad wavelength distribution results in a trade-off between a decrease in the diffraction efficiency and an increase in neutron flux. Thus, an intermediate thickness was adopted for the beamsplitter (G$_1$) and analyzer grating (G$_3$). In the first attempt, target grating thicknesses of about $d_{1,3}=25\,\mu\text{m}$ for $\text{G}_1$ and G$_3$, and $d_2=50\,\mu\text{m}$ for $\text{G}_2$, were chosen'. Note that a choice $d_2$ for G$_2$ is also crucial when it comes to contributions of higher diffraction orders ($|m|>1$) that might take place simultaneously in neutron diffraction. For $d\approx \Lambda^2/\lambda$, however, the second-order diffraction is substantially suppressed, yielding 40 to 50 $\mu\text{m}$ for this setup.

Finally, $\text{G}_1$ and $\text{G}_2$ are mounted on super-planar aluminum plates, while $\text{G}_3$ is mounted on a high-precision 6-axis piezoelectric stage (Physik Instrumente P-562.6CD; total linear translation path along $x,y,z$: 200 $\mu\text{m}$ each; total angle $\alpha_x,\alpha_y,\alpha_z$: 1 mrad each with a resolution of 0.1 $\mu\text{rad}$). The latter allows for precise mutual adjustment with respect to G$_1$ and G$_2$. Additionally, a phase difference $\Delta\chi$ between the interfering beams can be applied by shifting G$_3$ along the grating vector.

% For the second campaign three new nDPC-gratings $\text{C}_1,\text{C}_2,\text{C}_3$ were prepared employing the same procedure but this time with identical target thicknesses of $D_{1,2,3}=50\,\mu\text{m}$. This was triggered by the fact that using a specific part of the incoming beam which is redirected to the interferometer by a rather narrow bandpass supermirror, the spectra has a mean wavelength of $\overline{\lambda_\text{N}}\approx 4.4\m\text{nm}$ and a width of $\Delta\lambda/\lambda \lesssim 0.2$, which is very similar to the spectra investigated in detail recently  \cite{Neulinger-nima24a}, and thus a reduction of thickness can be avoided.

\section{Experimental results: LIGHT- and NEUTRON-OPTICAL CHARACTERISTICS OF THE INTERFEROMETER NDPC-GRATINGS}
Light optical characterization was done in the holography lab at the University of Vienna where a holographic setup together with alignment tools for successively recorded gratings is available. 
This characterization of the gratings involved:
\begin{enumerate}
\item the time dependence of the first-order diffraction efficiency $\eta_1(t;\Theta\approx\Theta_B)$ during recording,
\item the angular dependence of the first-order diffraction efficiency $\eta_{1;\text{L}}(\Theta)$ (rocking curve) after recording,
\item the positional dependence of the rocking curve across the hologram area.
\end{enumerate}
Whenever the evaluation of the light optical characteristic parameters, e.g., thickness or refractive-index modulation did not meet the requirements, such gratings were discarded and other ones were examined for use. From our previous systematic approach  \cite{Falmbigl-spie25} we know that satisfactory gratings judged from our diffraction-efficiency measurements are also usable as neutron-optical diffractive elements.
For our characterization of recorded gratings at neutron wavelengths, the rocking curves for neutrons at $-3,\ldots +3$ diffraction orders were measured.
\subsection{Experimental results: light optical properties of the gratings}
\begin{figure}[ht]
   \begin{center}
   \begin{tabular}{c} %% tabular useful for creating an array of images 
    \includegraphics[width=.95\columnwidth]{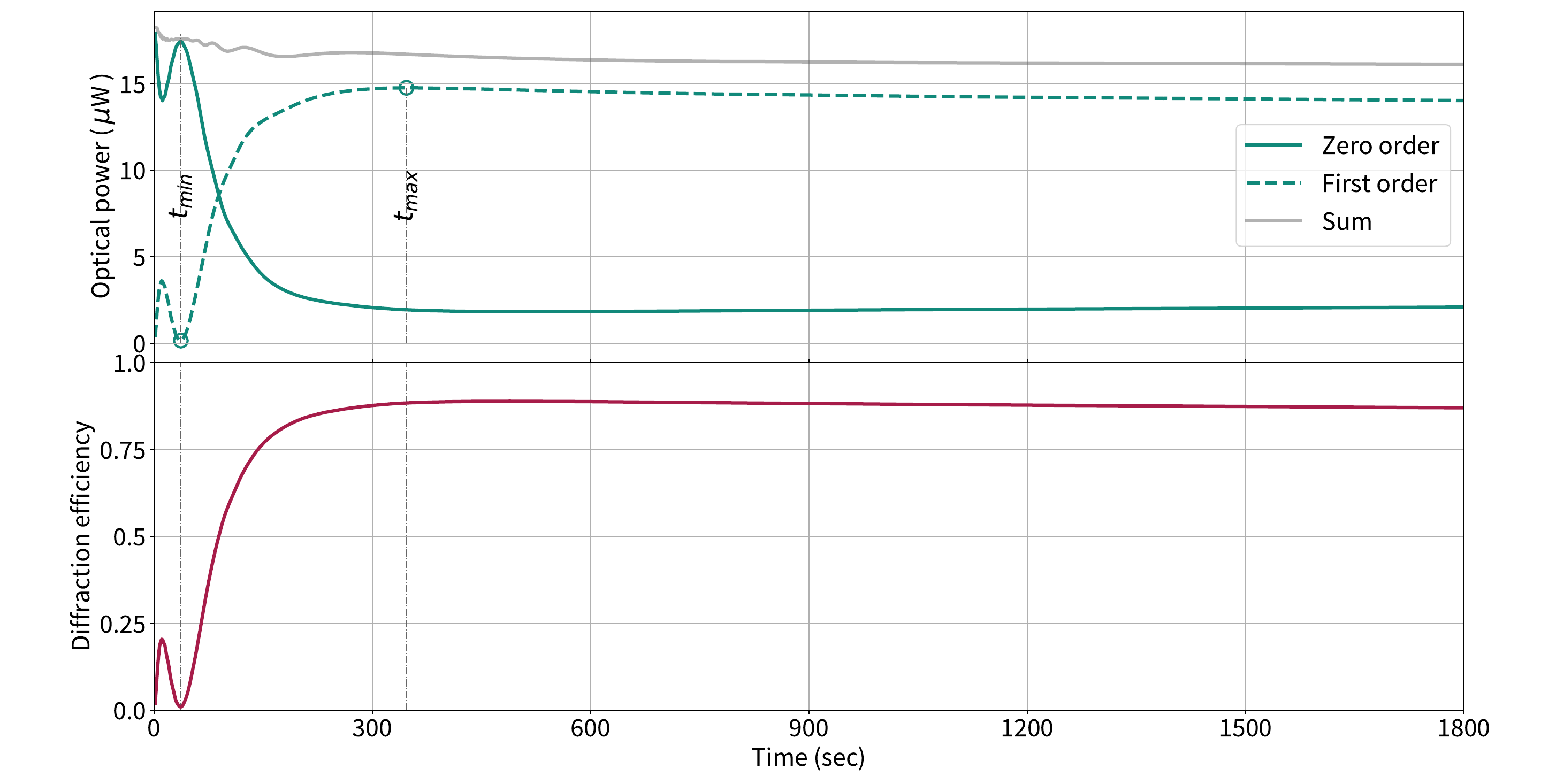} 
   \end{tabular}
   \end{center}
   \caption[G2light] 
   { \label{fig:time_dep} Temporal grating buildup dynamics during holographic recording for $\text{G}_2$ during the recording time (30 min) monitored using a red He-Ne laser ($\lambda_R=633\,\text{nm}$.) 
   \textit{Top panel}: Time evolution of zeroth- and first-order optical powers and their sum. \textit{Bottom panel} : Time evolution of the calculated diffraction efficiency from the data above, see  \cite{Ackermann-M26}.}
   \end{figure}

Light optical characterization yields a grating period of $\Lambda=(505.7\pm 1.4)\,\text{nm}$ for each of the gratings. Unslanted gratings were recorded by using an $s$-polarized beam of a diode-pumped solid state laser (Coherent Genesis SLM) at a wavelength of $514\,\text{nm}$. The coherently split beams of equal intensities ($100\,\text{mW/cm}^2$ each) interfered at the sample position at an angle of $2\Theta_B = 61.08^\circ$.
The time dependence of the first order diffraction efficiency, $\eta_1=I_1/(I_1+I_0)$ in close vicinity of the Bragg angle was monitored using light at a wavelength of $633\,\text{nm}$ (to which the unexposed material is photo-insensitive). Here, $I_{0,1}$ are the zero and first order diffracted intensities. The result for $\text{G}_2$ is depicted in Fig. \ref{fig:time_dep}.
The time dependence clearly shows an initial time interval with non-monotonic behaviour, followed by an increase of the first-order diffraction efficiency from a local minimum at $t_\text{min}\approx 36\,\text{sec}$ (lower panel) until the maximum value of about $89\,\%$ is obtained at $t_\text{max}\approx 350\,\text{sec}$ with a slight decrease until the end of the recording time. It is well known that the diffraction efficiency in the two wave-coupling model is a periodic function of the grating strength $\nu$ \cite{Kogelnik-atj69}:
\begin{equation}
	\label{eq:nu_eta}
	\eta_1(\Theta_B)=\sin^2(\nu), \quad \nu=\frac{n_1 d\pi}{\lambda\sqrt{1-\left(\frac{\lambda}{2n_0\Lambda}\right)^2}}.
\end{equation}
Here, $n_0$ is the average refractive index and $n_1$ the refractive-index modulation with $n(x)=n_0+n_1\cos(2\pi x/\Lambda)$ a volume phase grating.

\begin{figure}[thb]
   \begin{center}
   \begin{tabular}{c} %% tabular useful for creating an array of images 
    \includegraphics[width=.95\columnwidth]{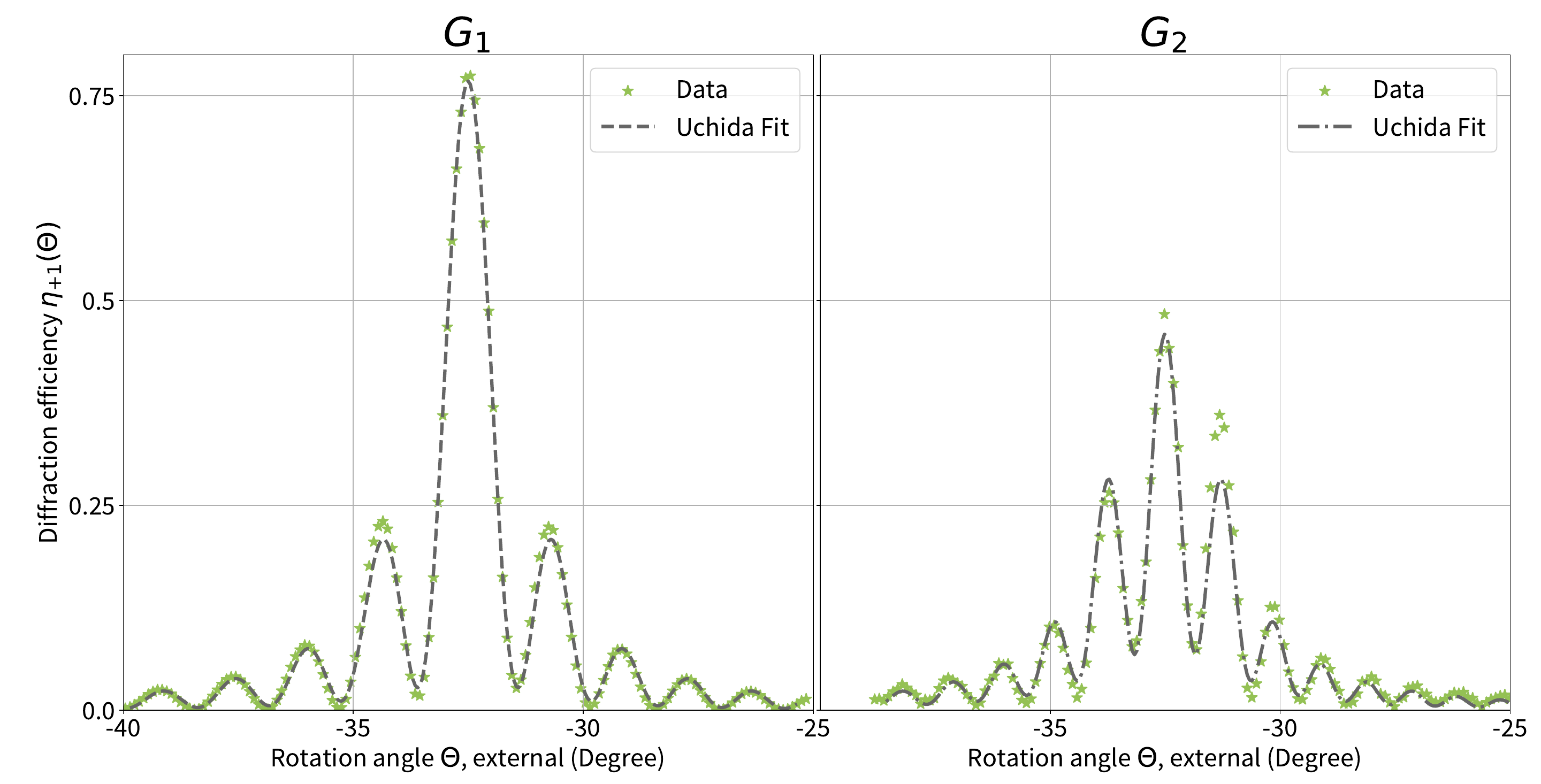}
   \end{tabular}
   \end{center}
   \caption[Glight] 
   { \label{fig:Glight} Angular dependence of the diffraction efficiency for interferometer gratings $\text{G}_1$ (left) and $\text{G}_2$ (right) probed with green laser light. Extracted grating parameters $n_1, d, L$ are summarized in Table \ref{tab:1}.}
   \end{figure}

% We argue, that the red laser \yco{was not at the perfect Bragg angle $\Theta_B$} and thus the maximum does not reach 100 \%. The decrease is explained by passing the regime to overmodulation for light, i.e., $\nu>\pi/2$. Similar curves were measured for the beamsplitter gratings $\text{G}_1$ and $\text{G}_3$, however, with lower maximum diffraction efficiencies due to their smaller thicknesses. 
% \yco{Q:WHY SO? --->
% mF answer: this was actually due to the fact that we aligned it using a thin Bayfol and then a 30-micron thick nDPC. Thus we did not recognize a small angular deviation, their DE was nearly 100\%. Using the 50-micron thick nDPC it showed up. Should we skip it?}
% 
% \yco{Q: WHY DO YOU SAY THAT DESPITE THE FACT THAT \ETA IS LOWER THAN UNITY? CAN YOU SHOW THAT FROM ITS ANGULAR PLOT AT 633NM AS SIMILAR TO AT 543.5 NM SHOWN BELOW?--> mF answer: no, we do not have the rocking curve using the red laser.}
The angular dependence of the diffraction efficiency $\eta_{+1}(\Theta)$ for each grating after recording was probed using a green He-Ne laser ($\lambda_G=543.5\,\text{nm}$). The experimental data are shown in Fig. \ref{fig:Glight} together with a fit to Uchida's modified two wave-coupling approach \cite{Uchida-josa73}. The latter accounts for an exponential attenuation of the refractive-index modulation along the grating depth $z$ as $n_1(z)=n_1(0)\exp(-z/L)$ with $L$ the decay length.
   %%%NOW: TABLE 1 light optical data
   \begin{table}[ht]
% \begin{center}
\renewcommand{\arraystretch}{1.3} 
% \begin{minipage}{\columnwidth}
\begin{center}
\begin{tabular}{|c|ccc|}\hline
\textbf{}&$d\,(\mu\text{m})$ & $\overline{n}_1$ & $L\,(\mu\text{m})$\\\hline 
$\text{G}_1$ & $(33.87 \pm 0.07)$  & $(0.991 \pm 0.003)\times 10^{-2}$ & $(50.1 \pm 1.9)$\\ 
$\text{G}_2$ & $(48.0 \pm 0.2)$  & $(0.808 \pm 0.005)\times 10^{-2}$ & $(40.5 \pm 1.4)$\\ 
$\text{G}_3$ & $(33.31 \pm 0.05)$   & $(0.990 \pm 0.002)\times 10^{-2}$ & $(57.6 \pm 2.0)$\\ \hline
\end{tabular}
\end{center}
% \end{minipage}
\caption[LIGHTO]{Summary of light optical grating parameters of the gratings employed in the VCN interferometer.\label{tab:1}}
\end{table}
   
\begin{figure}[hbt]
   \begin{center}
   \begin{tabular}{|c|c|} %% tabular useful for creating an array of images 
   \hline
%     \frame{
    \includegraphics[width=.65\columnwidth]{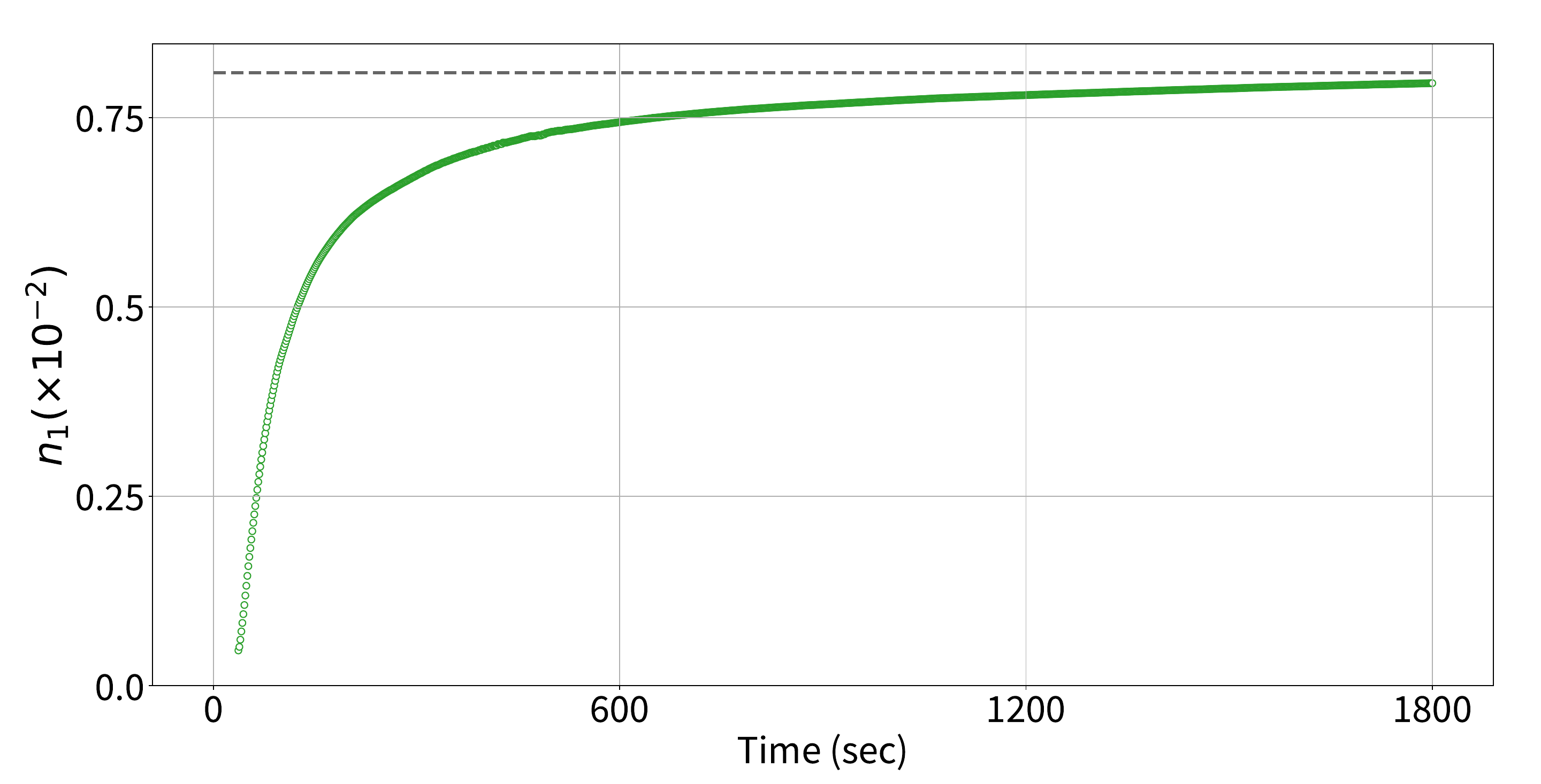} & \includegraphics[width=4.15cm]{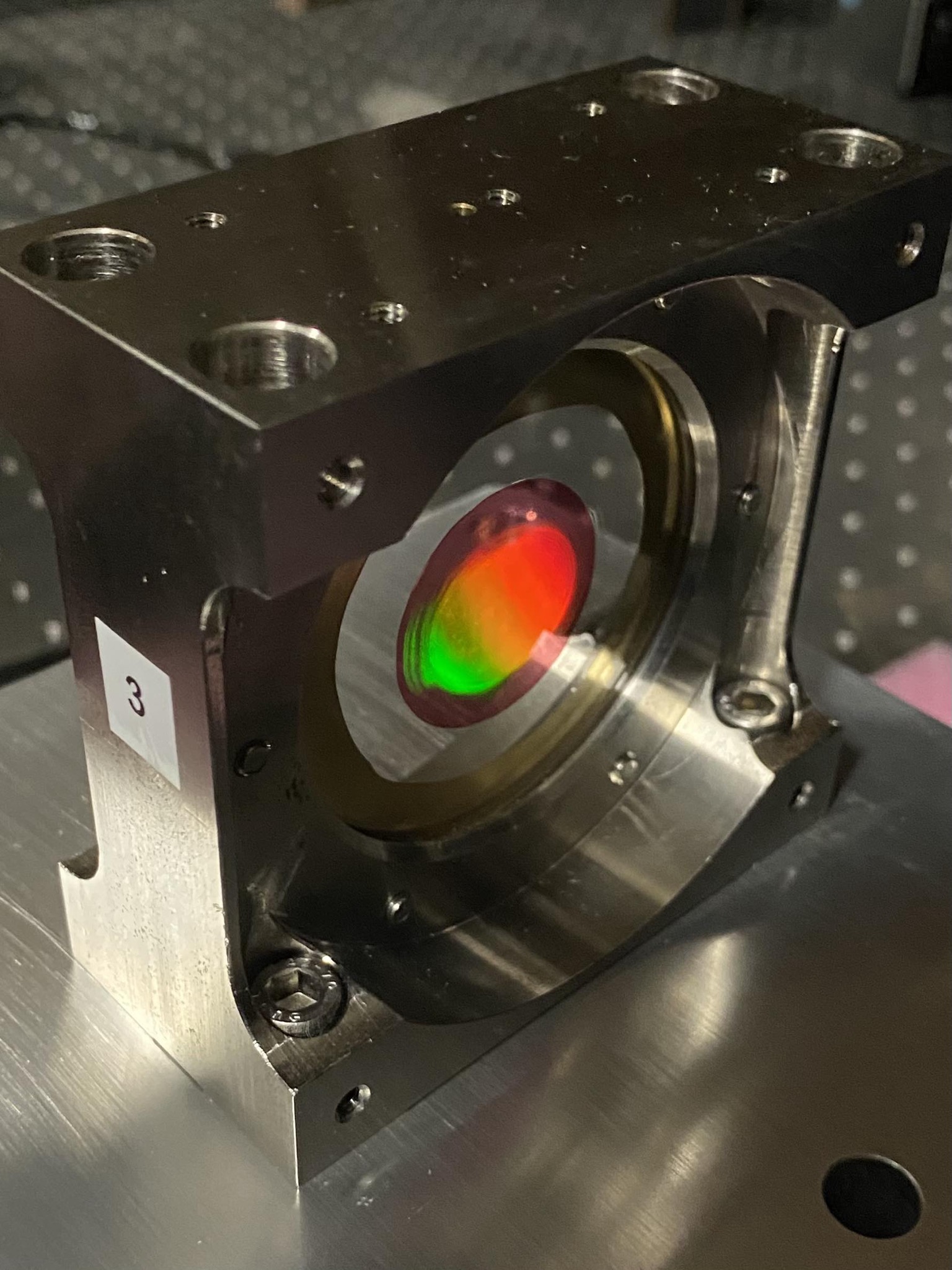}
%     }
    \\\hline
   \end{tabular}
   \end{center}
   \caption[Glight] 
   { \label{fig:n1} \textit{Left panel:} Temporal evolution of $n_1$ until saturation ($\Delta n_\text{sat}=0.81\times 10^{-2}$) for $\text{G}_2$.
%    \textit{Center panel: } $\eta(dn_1,\lambda)$ at the corresponding Bragg angles, see Eq. (\ref{eq:nu_eta}). Combinations which were realized for the gratings $\text{G}_{1,2,3}$ at the readout wavelengths are marked. 
   \textit{Right panel: } a photograph of $\text{G}_3$ in its sample holder \cite{Ackermann-M26}.}
   \end{figure}

   %%%%
Results for $\text{G}_3$ are very similar to those of $\text{G}_1$. The rocking curves and related fitting parameters clearly show that \textbf{both} gratings are overmodulated at 543.5-nm green light. Using the results from the fits to the rocking curve at $\lambda_G=543.5\,\text{nm}$ and the temporal evolution of $\eta(t)$ monitored by red light (see Fig. \,\ref{fig:time_dep}) the growth of the refractive-index modulation $n_1(t)$ during recording can be evaluated, as shown in Fig.\,\ref{fig:n1} for grating $\text{G}_2$.
% The wavelength dependence of the diffraction efficiency in Bragg condition upon increasing $n_1$ is also depicted in Fig. \,\ref{fig:n1}.
Furthermore, it was found that upon post-exposure of the recorded grating to a white light source (LED), $n_1$ slightly increases until a steady state is reached (over 2-day illumination).
 
The final optical characterization probes the diffraction efficiency across the hologram area (diameter $\approx 25\,\text{mm}$). Lateral shifts (horizontal and vertical) in steps of $5\,\text{mm}$ were performed. Fits to the corresponding rocking curves showed that the refractive-index modulation within an area of $20\times 20\,\text{mm}^2$ has a maximum relative standard deviation of less than 6\% (for $\text{G}_2$).
% % % % % % % % % % % % % % % % % % % d
\subsection{Experimental results: neutron-optical properties of the gratings}
The first step, before assembly of the three gratings into a VCN interferometer, was the neutron-optical characterization of each of the three gratings.
% Before assembling the three gratings to form an interferometer for VCN, we 
% characterized the gratings with respect to their transmission for the actual neutron beam with finite wavelength distribution. This is of some importance, as the neutron flux in collimated beams is strongly reduced and the number of neutrons reaching the detector is limited. As we did not intend to do a precision experiment,
% measured  with neutrons. 

To determine the attenuation of the gratings, the number of neutrons reaching the detector was counted 1) without any sample in place, and 2) for each of the gratings which were approximately $5^\circ$ off normal incidence to ensure that Bragg diffraction does not occur. The observed attenuation is due to  neutron absorption and incoherent/coherent scattering. It mainly originates from an nDPC grating material, the glass substrates of $3\,\text{mm}$ thickness do not  contribute significantly. 
% \yco{mF: I have now used the definition of an attenuation factor $\cal A$ and the linear attenuation coefficient $\mu$ as given in the Boothroyd book}
The linear attenuation coefficient $\mu$ for nDPC material amounts to $\mu=(8.4\pm 0.5)\,\text{mm}^{-1}$. For the three gratings a net attenuation factor $\cal A$ of 37\% can be estimated.  
% three nDPC samples of equal thickness of $50\,\mu\text{m}$ it will be reduced to even 28\%, which is a significant loss.
% Knowing the thicknesses of each grating and the silica glass plates we could get the total cross section, i.e., 

\begin{figure}[ht]
   \begin{center}
   \begin{tabular}{c} %% tabular useful for creating an array of images 
    \includegraphics[width=.95\columnwidth]{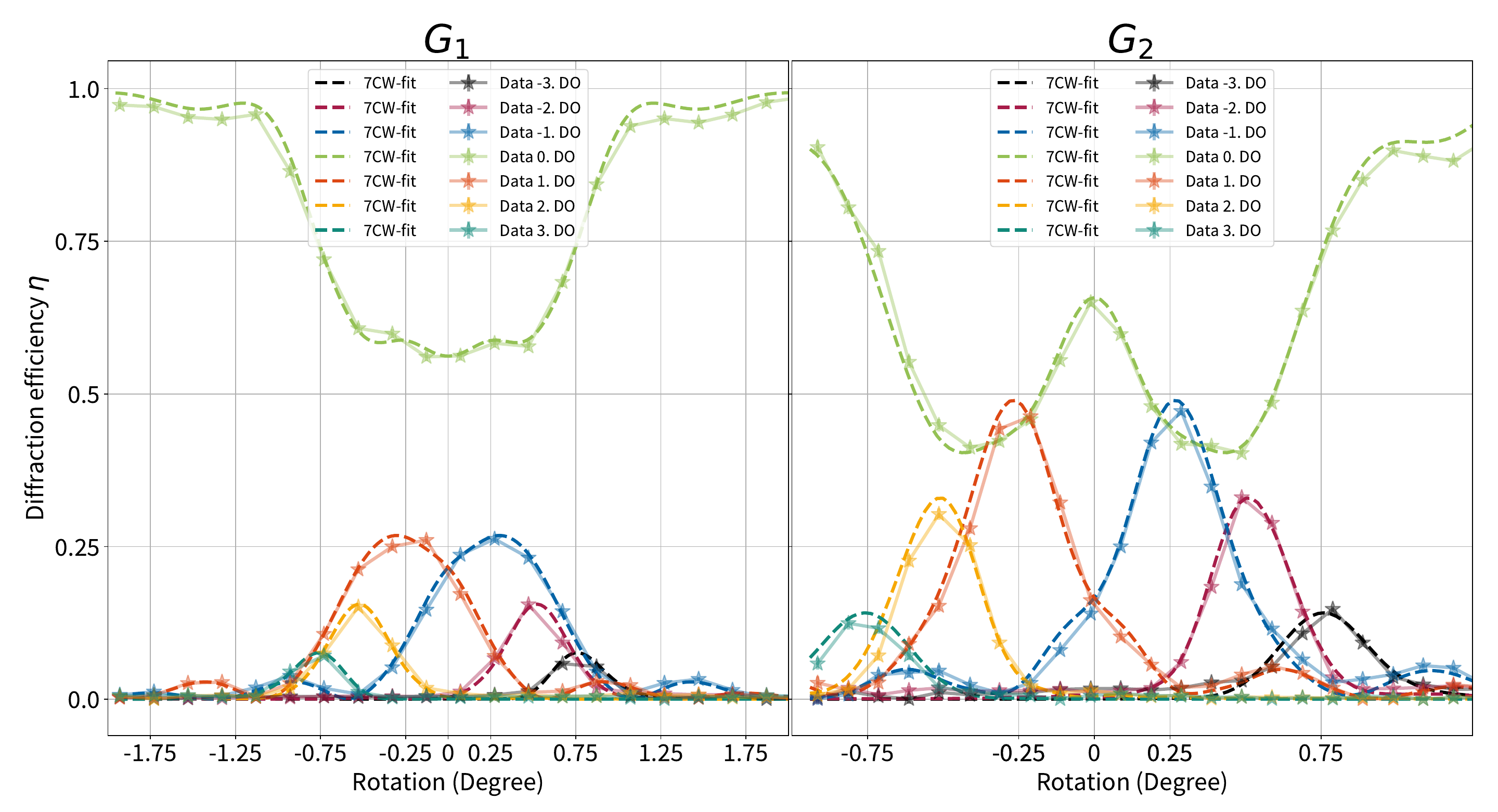}
   \end{tabular}
   \end{center}
   \caption[G1neutrons] 
   { \label{fig:Gneutron} Angular dependence of the diffraction efficiency for interferometer gratings $\text{G}_1$ (left) and $\text{G}_2$ (right) employing VCN. Data points including error bars are shown as $\star$, DO refers to diffraction order. Dashed lines are simultaneous fits to the data using a multiwave-coupling (7CW) theory  \cite{Klepp-jpcs16,Fally-oe21,Hadden-PhD24}. Extracted grating parameters are summarized in Table \ref{tab:1}.}
   \end{figure}

The individual diffraction efficiencies of G$_1$ and G$_3$ determine, among other important parameters, the visibility of the VCN interferometer given by $v=(I_\text{max}-I_\text{min})/(I_\text{max}+I_\text{min})$, where $I_\text{min,max}$ are maxima and minima of $I_0,I_H$ from Eq.\,(\ref{eq:1}). 
% Before assembling the gratings to constitute the VCN interferometer, we measured the angular dependence of the diffraction efficiency for each of them separately. We installed a collimation system 
Before assembly of the gratings into the VCN interferometer, the angular dependence of the diffraction efficiency was measured separately for each of them. A collimation system was installed at a distance of $1210\,\text{mm}$ between two slits with a horizontal width of $1\,\text{mm}$ for entrance as well as exit slits, which leads to a collimation of $\Delta\theta\approx 2\,\text{mrad}$. The wavelength distribution had an average wavelength of $\overline{\lambda}_N=4.5\,\text{nm}$ and $\Delta\lambda_\text{N}/\overline{\lambda}_\text{N}=0.1$ \cite{Neulinger-nima24a}. The experimental results of $\eta(\Theta)$ for $\text{G}_{1,2}$ are shown in Fig.\,\ref{fig:Gneutron}.
As expected, several diffraction orders ($\pm 3,\pm 2, \pm 1, 0$) were observed. For the VCN interferometer the zero and first orders at the Bragg angle $\Theta_{B;1}\approx 0.25^\circ$ were selected. As discussed in section \ref{sec:2}, for a grating with thickness $d\approx \Lambda^2/\lambda$ all other orders are negligible (see Fig. \ref{fig:Gneutron}, right panel.) From the measured angular dependence 
we evaluated the actual wavelength distribution ($\overline{\lambda}_\text{N}=(4.34\pm 0.05)\,\text{nm}$ and $\Delta\lambda_\text{N}=(1.7\pm 0.04)\,\text{nm}$) and the neutron-optical grating parameters (thickness $d$, and modulation amplitude $\Delta[b_c\rho]_{1,2,3}$ of the coherent scattering length density (SLD)) by fitting a multiwave-coupling model to the data; details can be found elsewhere \cite{Klepp-jpcs16,Hadden-apl24,Hadden-PhD24}.
%%%%%  evaluate the actual the wavelength distribution (three parameters) as well as the grating parameters (thickness $d$, and modulation amplitude $\Delta[b_c\rho]_{1,2,3}$ of the coherent scattering density (SLD)) a fit to the data using a multiwave-coupling model was performed, see e.g. Refs.  \cite{Klepp-jpcs16,Hadden-apl24,Hadden-PhD24}. The fitted EMG distribution $S(\lambda_\text{N})$ yields the parameters $\ell=(2.20\pm 0.18)\,\text{nm}^{-1}, \mu=(3.89\pm 0.03)\,\text{nm}, \sigma=(0.63\pm 0.04)\,\text{nm}$, i.e. an average wavelength of $\overline{\lambda}_\text{N}=(4.34\pm 0.05)\,\text{nm}$ and a FWHM of $(1.7\pm 0.04)\,\text{nm}$.%%%%from MonteCarlo sims.%%%%%%
Table \ref{tab:2} summarizes the neutron-optical parameters for the three gratings used in the VCN interferometer.
From Eqs. (\ref{eq:I0H}) the (theoretical) visibilities for the $I_0\, (I_H)$ beam can be deduced to be $0.72\, (1.00)$. 

\begin{table}[hbt]
\renewcommand{\arraystretch}{1.3} 
%%%%%%%%%HERE come the NEUTRON DATA

\begin{minipage}{2\columnwidth/3}
\begin{center}
\begin{tabular}{|c|cccc|}\hline
\textbf{(a)} &$d \,(\mu\text{m})$ & $\Delta[b_c\rho]_1 \,(\mu\text{m}^{-2})$ & $\Delta[b_c\rho]_2 \,(\mu\text{m}^{-2})$ & $\Delta[b_c\rho]_3 \,(\mu\text{m}^{-2})$\\\hline 
$\text{G}_1$ & $(35.5 \pm 0.2)$  & $(9.95 \pm 0.09)$ & $-(4.64 \pm 0.07)$ & $(2.04 \pm 0.08)$\\ 
$\text{G}_2$ & $(49.3 \pm 0.1)$  & $(10.23 \pm 0.06)$ & $-(5.09 \pm 0.04)$ & $(2.64 \pm 0.05)$ \\ 
$\text{G}_3$ & $(34.5 \pm 0.1)$   & $(9.33 \pm 0.06)$ & $-(4.48 \pm 0.05)$ &$ (1.99 \pm 0.06)$\\ \hline
\end{tabular}
\end{center}
\end{minipage}
\begin{minipage}{\columnwidth/3-5mm}
\begin{tabular}{|c|cc|}
\hline
 \textbf{(b)}& first order & zero order \\\hline
$\text{G}_1$ & $\ej{1}{1}=27\%$ &$\ej{1}{0}=59\%$\\
$\text{G}_2$ & $\ej{2}{1}=49\%$ & $\ej{2}{0}=44\%$\\
$\text{G}_3$ & $\ej{3}{1}=24\%$ & $\ej{3}{0}=60\%$\\ \hline
\end{tabular}
\end{minipage}
\vspace*{1ex}%%%%%%%%%HERE come the NEUTRON DATA

\caption[summary]{\label{tab:2} 
% \textbf{()} Summary of light optical grating parameters of the gratings employed in the VCN interferometer.\\
\textbf{(a)} Summary of the neutron-optical grating parameters for G$_{1,2,3}$ for an EMG wavelength distribution.\\
\textbf{(b)} Experimentally determined diffraction efficiencies for first and zero orders for each of the gratings of the VCN interferometer. See text for meaning of the symbols.}
\end{table}
\section{THE INTERFEROMETER at the PF2-VCN beamline}
\begin{figure}[htb]
   \begin{center}
%    \begin{tabular}{c} %% tabular useful for creating an array of images 
    \includegraphics[width=.9\columnwidth,trim=0 0 0 90,clip]{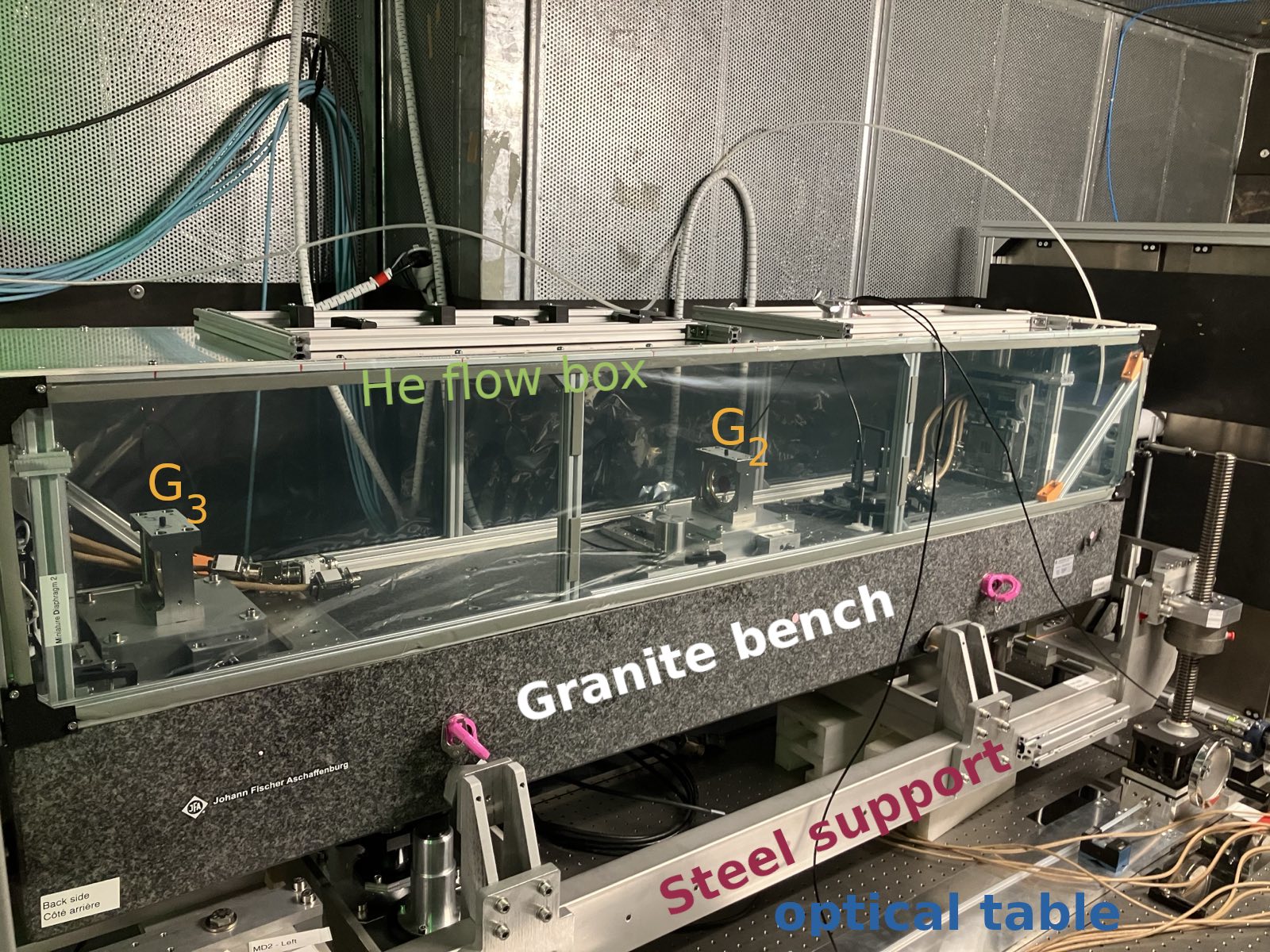}\\ %& \includegraphics[width=.5\columnwidth]{image-2025-10-10-180512.png}
   %%%{signal-2025-10-10-101227.jpeg}
%    \end{tabular}
   \includegraphics[width=\columnwidth,trim=0 25 0 150,clip]{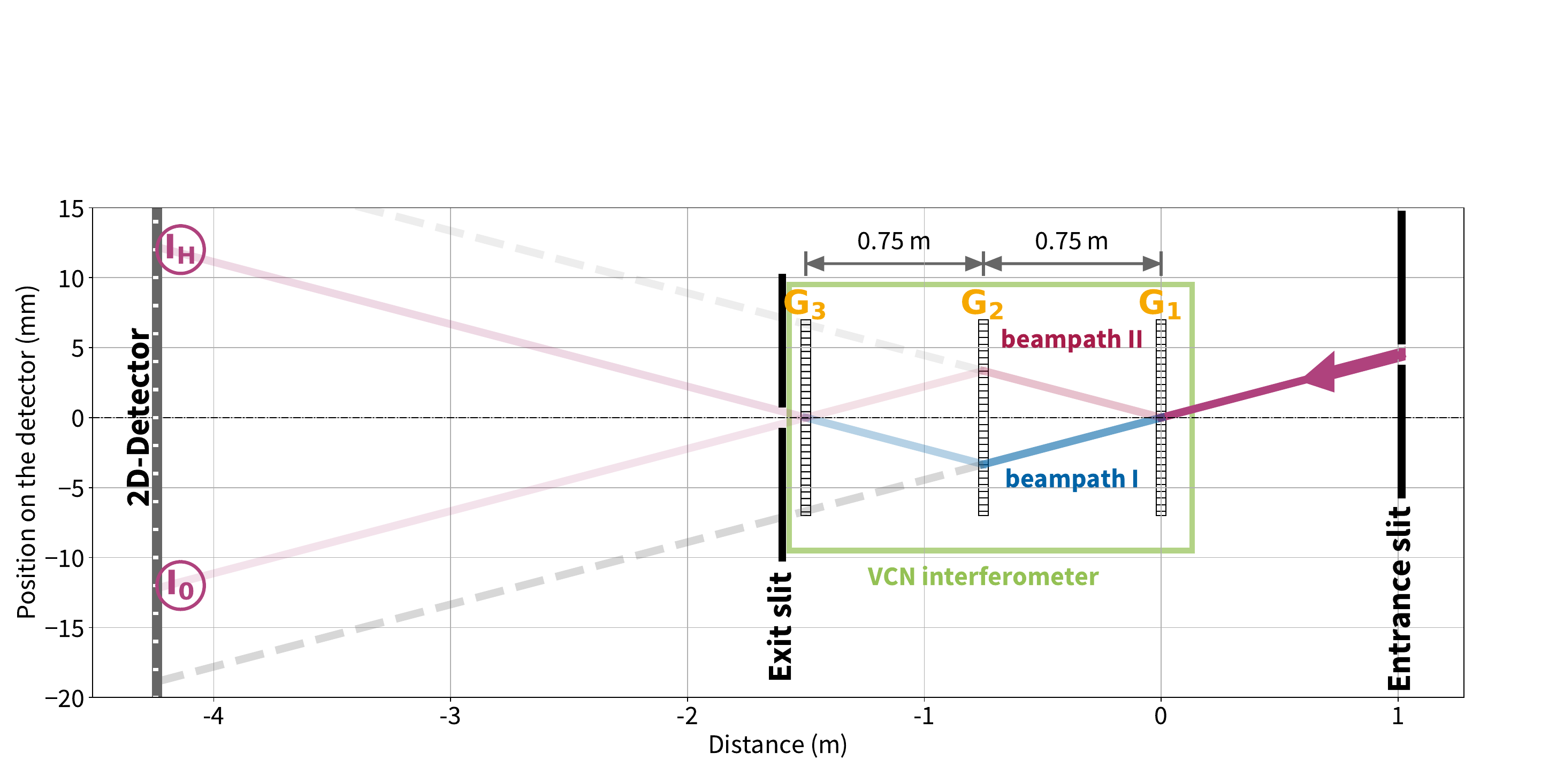}%{image-2025-10-10-180512.png}
   \end{center}
   \caption[VCN] 
   { \label{fig:VCN} \textit{Top panel: } Photograph of the VCN-interferometer (gratings on granite bench; G$_1$ hidden behind a slit on the right) on the steel support placed on top of an optical table. Air cushions allow to move the entire VCN-interferometer on the optical table. Neutrons enter from the right side, detector on the left side (not visible).
   \textit{Bottom panel: } Schematic of the geometry for the VCN-interferometer. The position of the gratings, slits, the beams and the detector are shown.
%    \textcolor{uorange}{mF will prepare a new version for this proceedings, but needs time.}
   }
   \end{figure}

Figure\,\ref{fig:VCN} shows a photograph of the setup during beamtimes 3-14-455 \cite{Klepp-InstitutLaue-LangevinILL25} and DIR-407 \cite{Klepp-InstitutLaue-LangevinILL25a} at the PF2-VCN beamline (Institut Laue-Langevin, Grenoble, France) as well as a schematic of the VCN interferometer geometry, i.e., beams, slits, gratings and the detector.
After having characterized the neutron-optical properties of the three gratings they were placed equidistantly ($75\,\text{cm}$; total interferometer length: $150\,\text{cm}$) on the ultra-flat granite bench (see Fig.\,\ref{fig:VCN}, top panel) and aligned with respect to their yaw angles using an autocollimation system (prealignment of positions and specifications of the flatness were provided by \textit{Fa. Johann Fischer, Aschaffenburg}). The granite bench was placed on a sophisticated support (designed at the ILL) that could glide on top of an optical table. Further, passive damping
% on top of an optical table. Passive damping 
elements were deployed to reduce vibrations at a broad range of frequencies. Remaining vibrations were thoroughly monitored. The PF2-VCN beamline is situated in a container-like housing where 
% temperature stabilization could be established. 
the temperature could be actively stabilized.
Furthermore, the VCN interferometer was covered with a box which was filled with He gas at a low flow rate to prevent scattering of the neutrons from air molecules. 

The entire VCN interferometer was moved (by using air-cushions) to optimal positions, i.e., 1) the position of first order Bragg diffraction for each of the three gratings with respect to the incident neutron beam and 2) positions at the centre of the gratings $\text{G}_1$ and G$_3$. Due to the fact that $\text{G}_2$ serves as mirror for \textbf{two} beams, the latter are off-centred by about $\pm 4\,\text{mm}$ to the left and the right (see Fig.\,\ref{fig:VCN}, bottom panel). Therefore, $\text{G}_2$ is required to be a perfectly periodic grating over a larger area, say of at least $1 - 1.5\,$ cm. 
% The entire interferometer had to be moved, lifted up, tilted etc. to find its best position. 
Collimation was provided by motorized slits well in front of the grating $\text{G}_1$ and behind grating $\text{G}_3$ (distance longer than $ 2.5\,\text{m} $). 
Additional slits were inserted near $\text{G}_2 $ to block the parasitic beams as discussed above (see Fig.\,\ref{fig:VCN}). A two-dimensional neutron detector was placed at a further distance of $2.2\,\text{m}$ downstream of $\text{G}_3$.

% Tight control of the temperature and vibration stability is especially important due to the long measurement time, which is required due to the low neutron count rates. 
Owing to the low neutron count rates - and the resulting long measurement times - precise temperature control and vibrational stability are essential.
% Control of the temperature and vibration stability, which do have a strong influence on the phase stability of the interference pattern, 
% are important also in ragard to low neutron count rates due to which very long measurement time is required. 
% Unfortunately, we encountered that the thermal power was at a lower level than the maximum power available in the first reactor cycle. 
A trade-off between the phase stability and sufficient neutron counting statistics has to be made.
% Thus, any improvement of the (ideal) visibility and/or reduction of losses is necessary and very welcome. Actually, setting up and aligning the VCN-interferometer during a 3-week campaign is a tour-de-force and very demanding.  
\section{Discussion and next steps: options for optimizing the VCN interferometer further}
Phase scans have shown that the instrument works in principle, i.e. oscillations can be sporadically observed but sufficient phase stability is not reached at the present level of development. Thus the instrument is not yet suitable for dedicated experiments. There are various options to advance the VCN interferometer's performance.

First, we will focus on the gratings' properties with higher diffraction efficiencies (for the beamsplitter gratings) and lower transmission losses. Diffraction efficiency at the Bragg angle can be estimated using the two wave-coupling theory, provided that the thickness and the grating spacing are properly chosen. One parameter to check the validity is the sum of the zeroth and the first-order diffraction efficiencies $\eta_0$ and $\eta_1$. When $\eta_0+\eta_1\approx 1$, the two wave theory is applicable and the first order diffraction efficiency for neutrons then can be approximated to 
\begin{equation}\label{eq:DE_N}
\eta_\text{1;N}=\sin^2\left(\frac{1}{2}\lambda_\text{N}d\,\Delta[b_c\rho]_1\right)
\end{equation}
It can be seen from Table\,\ref{tab:2}\textbf{(b)} that $\eta_1+\eta_0>0.9$ for G$_2$ and the above approximation is acceptable. 
It then follows that $\eta_\text{1;N}$ depends on three controllable parameters, the wavelength, the thickness, and the SLD modulation. 

% The use of a longer wavelength is an option 
It is possible to use a longer wavelength 
to increase neutron diffraction efficiency and Bragg angles. However, at existing neutron sources this option implies accepting an even lower incident neutron flux.

Increasing the thickness of gratings bears some difficulties in recording ideal gratings across a larger area but may be feasible. Therefore, for the second campaign four new nDPC-gratings were prepared employing the same procedure as described above but this time with identical target thicknesses of $50\,\mu\text{m}$. 
At PF2-VCN a similar wavelength distribution with a mean wavelength of $\overline{\lambda}_\text{N}\approx 4.4\,\text{nm}$ and a width of $\Delta\lambda_\text{N}/\lambda_\text{N} \simeq 0.1$ was realized  \cite{Neulinger-nima24a}. $\eta_\text{1;N}$ then is enhanced according to Eq.\,(\ref{eq:DE_N}). However, the net attenuation factor for three thick gratings reduces the total number of neutrons at the detector further to ${\cal A}\approx 28\%$. %%We have opted for this possibility, though.

The last option is to increase the SLD modulation. This could imply the use of other materials, e.g., polymer blends such as cyclic allylic sulfide based photopolymer  \cite{Galli-jps21} or other nanoparticles  \cite{Hadden-sr25}. Improvement of the recording conditions may also help to further increase the SLD modulation. 

The ultimate solution would be to increase the thickness and the SLD modulation with decreasing absorption and scattering losses. As the photopolymer in nDPC materials contains a large amount of hydrogen, it is obvious that incoherent scattering rather than absorption is the main reason for the massive loss. One way to mitigate this problem would be to make use of deuterated substances, a path that proved successful previously in binder-based photopolymer  \cite{Matull-zpb90}. However, the situation for nDPC materials might be different and could reduce not only attenuation but unfavorably also the SLD modulation: while nanodiamond particles have a positive SLD, \ce{^1H} has a negative one (cf. Ref.\, \cite{Sears-nn92}) but \ce{^2H} has nearly the same SLD as carbon. The SLD modulation of nDPC gratings foremost originates from photoinduced nanodiamond-rich regions and nanodiamond-poor regions. Thus deuteration may not improve the SLD modulation.

% Improvement from the instrument side can be done with respect to stabilization against vibrations and temperature changes as it has been pointed out that lack of vibration isolation and temperature stabilization reduce the contrast of the interference pattern drastically  \cite{Geppert-nima14,Lemmel-jac22} or even completely destroy it.

\section{SUMMARY}
% We report on test of an interferometer for very cold neutrons based on holographic gratings. 
A test of an interferometer for very cold neutrons based on holographic gratings is reported.
The gratings were characterized with respect to their light optical and neutron-optical performance. The first-order diffraction efficiency of the gratings $\text{G}_1 - \text{G}_2 - \text{G}_3$ for VCN with an average wavelength of $\overline{\lambda}_\text{N}=4.4\,\text{nm}$ was 27\% -- 49\% -- 24\%. This, in principle, allows visibilities of 100\%\,(72\%) for the $0\,(H)$ beams. A drawback of nDPC gratings is their considerable attenuation, even at thicknesses of a few ten micrometers. 
% The installation of the interferometer at the PF2-VCN was described and suggestions for improvement were discussed.
The installation of the interferometer at PF2-VCN was described, and possible improvements were discussed.

\acknowledgments % equivalent to \section*{ACKNOWLEDGMENTS}       
This research was funded in part by the Austrian Research Promotion Agency (FFG), Quantum-Austria NextPi, grant number FO999896034 and the European Union:
”NextGenerationEU”. EH is grateful to the Vienna Doctoral School in Physics (VDSP) for financial support, in particular a VDSP mobility fellowship. We particularly thank the associate director - science division of the ILL for granting us a \textit{Director's Discretion Time}. 
Our gratitude goes to the Bureau d’études, the Hall d’essais team, the SCI team, and Alexander Quirk for their invaluable technical work and support throughout the design and implementation at the ILL.

\bibliography{/home/fallym4/texmf/bibtex/bib/misc/isistr,/home/fallym4/texmf/bibtex/bib/misc/Data_DOI,/home/fallym4/texmf/bibtex/bib/misc/publications_url,/home/fallym4/texmf/bibtex/bib/misc/references24,/home/fallym4/texmf/bibtex/bib/misc/phd,/home/fallym4/texmf/bibtex/bib/misc/SPIE26} % bibliography data in report.bib
% % \bibliographystyle{fallycv} % makes bibtex use spiebib.bst
% \bibliographystyle{spiebib} % makes bibtex use spiebib.bst

\end{document}